\documentclass[preprint,12pt]{article}

\usepackage{amssymb}
\usepackage{graphicx}

\begin{document}




\centerline{\bf
A note on threshold theorem of fault-tolerant quantum computation}
\vspace*{0.37truein}
\centerline{
\footnotesize Min Liang and Li Yang} \vspace*{0.015truein}
\centerline{\footnotesize\it State Key Laboratory of Information
Security, Graduate University of Chinese Academy of Sciences,}
\centerline{\footnotesize\it Beijing 100049, China}

\vspace*{0.21truein}
%
%
%
%




Error-correction process has to be carried out periodically to prevent accumulation of errors in fault-tolerant quantum computation \cite{Gottesman09,Nielsen00,Aharonov99,Dennis01}. It is believed that the best choice to get maximum threshold value is carrying out an error-correction process after each quantum gate operation \cite{Nielsen00}. Result of this note shows that the optimal error-correction period depends on the value of $k$ which is the level number of concatenated quantum error-correction code (QECC).

We consider QECC $[m,1]$ which encodes $1$ qubit into $m$ qubits, and corrects any one qubit error \cite{Gottesman09,Steane96,Gottesman98,Knill05,Poulin06}. Denote the logical depth of its encoding circuit as  $\alpha$, that of its decoding circuit as $\beta$. If we concatenate this QECC $k$ times \cite{Knill96,Aliferis05}, then the logical depths of encoding and decoding circuit are $\alpha k$ and  $\beta k$, respectively.

	{\bf Definition 1:} Define the error-correction period $r$ as the maximum number of operations which act on single qubit between two successive error-corrctions in quantum circuit without fault-tolerant structure.

According to this definition, $r$ is the logical depth of algorithm executed within one error-correction period.

	{\bf Definition 2:} Fault-tolerant quantum circuit of one error-correction period consists of encoding circuit, decoding circuit, and a few steps of fault-tolerant computation between them.

Fault-tolerant quantum gates \{H, CNOT, S, Toffoli\} make up a universal set of gates for fault-tolerant quantum circuit. Based on the QECC we selected, some of these gates can be implemented transversally, and some of them have to be fault-tolerantly implemented via fault-tolerant measurement. If the gate can be implemented transversally, then the logical depth of its fault-tolerant quantum gates is 1, otherwise is bigger than 1. We denote the average logical depth of fault-tolerant quantum gates as $\delta$. Then, in fault-tolerant quantum circuit of each error-correction period, the logical depth is at most $\alpha k+\beta k+r\delta$.

    We denote $p$ as the probability of a failure in any one of the components used in the quantum circuit. In one error-correction period of fault-tolerant quantum circuit, the error probability on each physical qubit is $1-(1-p)^{\alpha k+\beta k+r\delta}\approx(\alpha k+\beta k+r\delta)p$.

  For a QECC correcting one qubit error, unrecoverable error occurs in the case that emerging more than two errors in one $m$-qubit block.

	In one error-correction period of fault-tolerant quantum circuit, the error probability on each physical qubit is  $\alpha k+\beta k+r\delta$, then the probability of an unrecoverable error occuring on one logical qubit of its upper level is $c((\alpha k+\beta k+r\delta)p)^2$, where $c$ is a constant. Note that we concatenate the QECC $k$ times. In one error-correction period, the probability of an unrecoverable error occuring on one logical qubit of the top level is $\frac{1}{c}\left[c(\alpha k+\beta k+r\delta)p\right]^{2^k}$.

	In the quantum circuit without fault-tolerant structure, the error probability on one qubit in one period is $rp$. Thus, we have a threshold condition
\begin{equation}
\frac{1}{c}\left[c(\alpha k+\beta k+r\delta)p\right]^{2^k}\leq rp,
\end{equation}
which means that, by using $k$-level concatenated codes \cite{Nielsen00,Knill96}, the error probability on one qubit in one period can be reduced from $rp$ to $\frac{1}{c}\left[c(\alpha k+\beta k+r\delta)p\right]^{2^k}$.
	
    This threshold condition is equivalent to
\begin{equation}
p\leq \frac{1}{c}\left(\frac{r}{(\alpha k+\beta k+r\delta)^{2^k}}\right)^{\frac{1}{2^k-1}},
\end{equation}
so we define the threshold value $p_{th}$\cite{Aliferis05,Steane03} as follows:
\begin{equation}\label{pth}
p_{th}=\frac{1}{c}\left(\frac{r}{(\alpha k+\beta k+r\delta)^{2^k}}\right)^{\frac{1}{2^k-1}}.
\end{equation}

For a given QECC, the parameters  $\alpha$, $\beta$, $\delta$, $c$ are deterministic. In fomula (\ref{pth}), $k$, $r$ and $p_{th}$  are unknown parameters. This fomula shows that, the threshold value $p_{th}$ depends on the level number $k$ and the error-correction period $r$. If we fix the value of $k$, then the value of $p_{th}$ depends only on the value of $r$, we can calculate the optimal error-correction period related to the threshold value.

For example, Steane code \cite{Steane96} is taken as the QECC. In this case, $m=7$ and $c=21$. The logical depth of encoding circuit is  $\alpha=4$, and the logical depth of decoding circuit is $\beta=10$. Let $\delta=2$, then we draw the following figures to show the relation of $p_{th}$ and $r$ when $k=1, 2, 3$ and $4$, respectively.

\begin{figure}[htb!]
\includegraphics[scale=0.65]{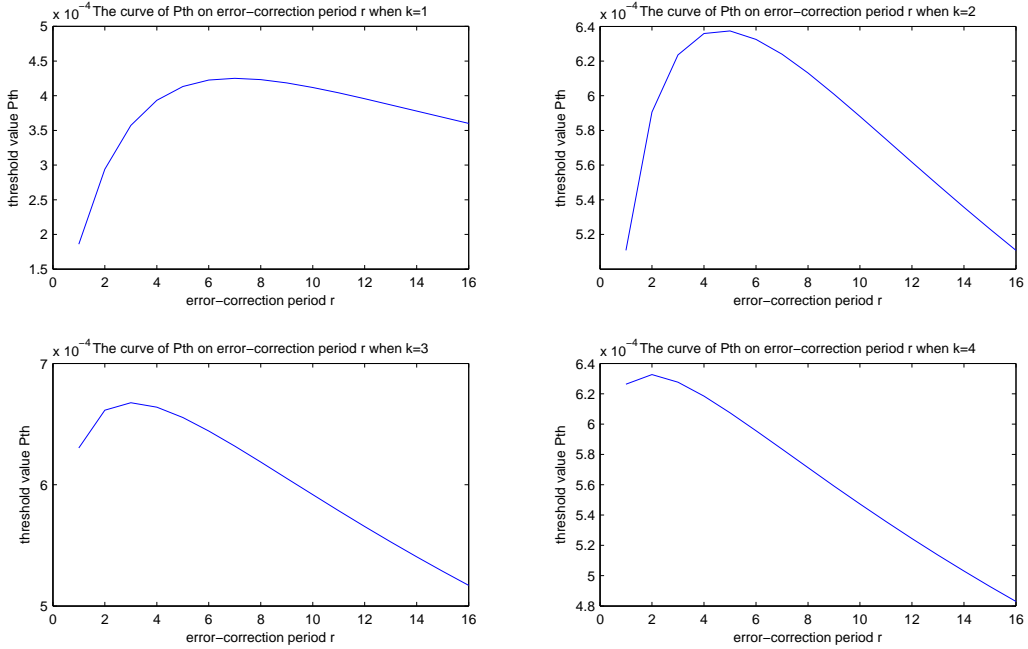}
\caption{The variation of $p_{th}$ with $r$. The Steane code is concatenated $k$ times, $k=1, 2, 3$ and $4$, respectively.}
\end{figure}

  It can be seen from these figures that, in order to improve the threshold value $p_{th}$, the error-correction period $r$ have to be adjusted according to the value of $k$.

Let $f(r,k)\triangleq\frac{1}{c}\left(\frac{r}{(\alpha k+\beta k+r\delta)^{2^k}}\right)^{\frac{1}{2^k-1}}$. By means of $\frac{\partial f(r,k)}{\partial r}=0$, we get $r=\frac{(\alpha+\beta)k}{\delta(2^k-1)}$. It can be verified that $\frac{\partial^2 f(r,k)}{\partial^2 r}<0$. Thus $f(r,k)$ arrives its maximum value at $r=\frac{(\alpha+\beta)k}{\delta(2^k-1)}$, which can be regarded as the optimal error-correction period.
This result shows that the optimal error-correction period $r$ depends on the value of $k$.







\end{document}